\documentclass[aps,pre,preprint]{revtex4-1}
\usepackage{lipsum}
\usepackage[T1]{fontenc}
\usepackage[utf8]{inputenc}
\usepackage{amssymb,bm}
\usepackage{amsmath,amsfonts,relsize}
\usepackage{graphicx,hyperref,xcolor}
\usepackage{lineno,setspace}
\usepackage{verbatim}
\begin{document}
\title{ Stationary properties of a Brownian gyrator with non-Markovian baths }
\author{Eduardo dos S. Nascimento}
 \email{edusantos18@esp.puc-rio.br}
 \affiliation{Dept. of Physics, PUC-Rio, Rio de Janeiro, Brazil}
\author{Welles A. M. Morgado}
 \email{welles@puc-rio.br}
 \affiliation{Dept. of Physics, PUC-Rio, Rio de Janeiro, Brazil}
 \affiliation{National Institute for Science and Technology - Complex Systems, Brazil}
\date{\today} 
\begin{abstract}
We investigate the stochastic behavior of a two-temperature Langevin system with 
non-Markovian thermal reservoirs. The model describes an overdamped Brownian particle in 
a quadratic potential and coupled to heat baths at different temperatures. 
The reservoirs are characterized by Gaussian white and colored noises and a 
dissipation memory kernel. The stationary states present non-trivial average 
rotational motion influenced by stochastic torques due to harmonic, friction 
and fluctuating thermal forces. However, the Markovian limit leads to a 
vanishing average torque produced by fluctuating thermal forces. 
We also study the effects of memory on the stochastic heat and the entropy 
production in the steady-state regime.
\end{abstract}
%
%
%
\maketitle
\section{Introduction} \label{Intro}

Many-body systems in thermodynamic equilibrium are characterized 
by distinct fundamental relationships between few macroscopic quantities, 
despite the huge number of microscopic degrees of freedom \cite{Callen1985,Huang1987}. 
Nevertheless, as length-scales and time-scales are reduced, non-trivial 
fluctuations play a relevant role in the microscopic and mesoscopic domains. 
In fact, the investigation of theses effects is a very important research topic for the understanding 
and technological developments of microscopic engines, in noisy environments
\cite{REIMANN200257,BroeckMeurs2004,Bustamante2005,Seifert2012,
Martnez2015,Ciliberto2017,FogedbyImparato2017,PietzonkaSeifert2018,Su2019,ManikandanKrishnamurthy2019}.

Thermal fluctuations on small-scale objects can be 
studied through a stochastic device proposed by 
Filliger and Reimann \cite{FilligerReimann2007}, which is called 
Brownian gyrator. This system consists of a Brownian particle trapped 
in a two-dimensional harmonic well and coupled to thermal baths at 
distinct temperatures. The reservoirs act along orthogonal spatial 
directions. The quadratic potential is supposed to be asymmetric, 
with principal axes different from the perpendicular directions associated with baths (see Fig.~\ref{2DLan}).

Brownian gyrators present interesting non-equilibrium properties which are described 
theoretically \cite{FilligerReimann2007,DotsenkoOshanin2013,MancoisWilkowski2018,CerasoliRondoni2018} 
and also in the context of experiments \cite{ArgunVolpe2016,ChiangChen2017}. 
The interplay between asymmetric potentials and orthogonal spatial directions 
associated with therml baths may give rise to emergent rotational motion, associated with the probability current~\cite{DotsenkoOshanin2013}. As the Brownian particle is coupled 
to heat reservoirs at different temperatures, the system behaves as a Brownian engine 
that exhibits systematic gyration (about the potential minimum) induced by thermal fluctuations.

From a theoretical point of view, Brownian gyrators are modeled as 
two-temperature Langevin dynamics in a quadratic potential \cite{FilligerReimann2007,DotsenkoOshanin2013,MancoisWilkowski2018,CerasoliRondoni2018}. In fact, 
many studies analyze the problem in terms of a Brownian particle with Markovian thermal reservoirs \cite{FilligerReimann2007,DotsenkoOshanin2013,MancoisWilkowski2018,CerasoliRondoni2018}. 
However, the presence of time-correlated stochastic forces can 
also lead to out-of-equilibrium properties \cite{PuglisiVillamaina2009,Villamaina2013}. 
Since the interesting fluctuating-induced behavior of Brownian gyrators are related to non-equilibrium 
effects, it could be interesting to consider the influence of memory on these stochastic systems. 

In this work, we consider a Brownian gyrator coupled to non-Markovian 
thermal reservoirs. The system is treated in the overdamped approximation, along the 
lines of discussions presented in  \cite{PuglisiVillamaina2009,VillamainaVulpiani2009,Villamaina2013,NascimentoMorgado2019}. 
The long-time limit leads to out-of-equilibrium stationary 
states as bath temperatures are distinct. The steady-state rotational motion is affected by 
torques produced by thermal fluctuating and friction forces, in addition to the torque exerted 
by harmonic forces. This is in contrast to the Markovian limit, 
which gives a vanishing mean torque due to thermal fluctuating forces \cite{FilligerReimann2007}. 
Memory also affects the stochastic heat exchanged with baths as well as the 
entropy production in the steady-state regime.

The paper is organized as follows. In Sec. \ref{2DLangevin}, 
we introduce the Brownian gyrator through a two-temperature Langevin system. 
The probability density for stationary states is presented in Sec. \ref{StatProb}. 
The average gyration properties are calculated in Sec. \ref{StoTor}. 
We discuss, in Sec. \ref{StoEner}, the stochastic heat exchanges. 
The entropy variation of steady-states is determined in Sec. \ref{StochEntroVar}. 
Finally, we present the conclusions in Sec. \ref{Con}.

\section{ Two-temperature Langevin dynamics } \label{2DLangevin}

\begin{figure}
 \centering
 \includegraphics[scale=0.5]{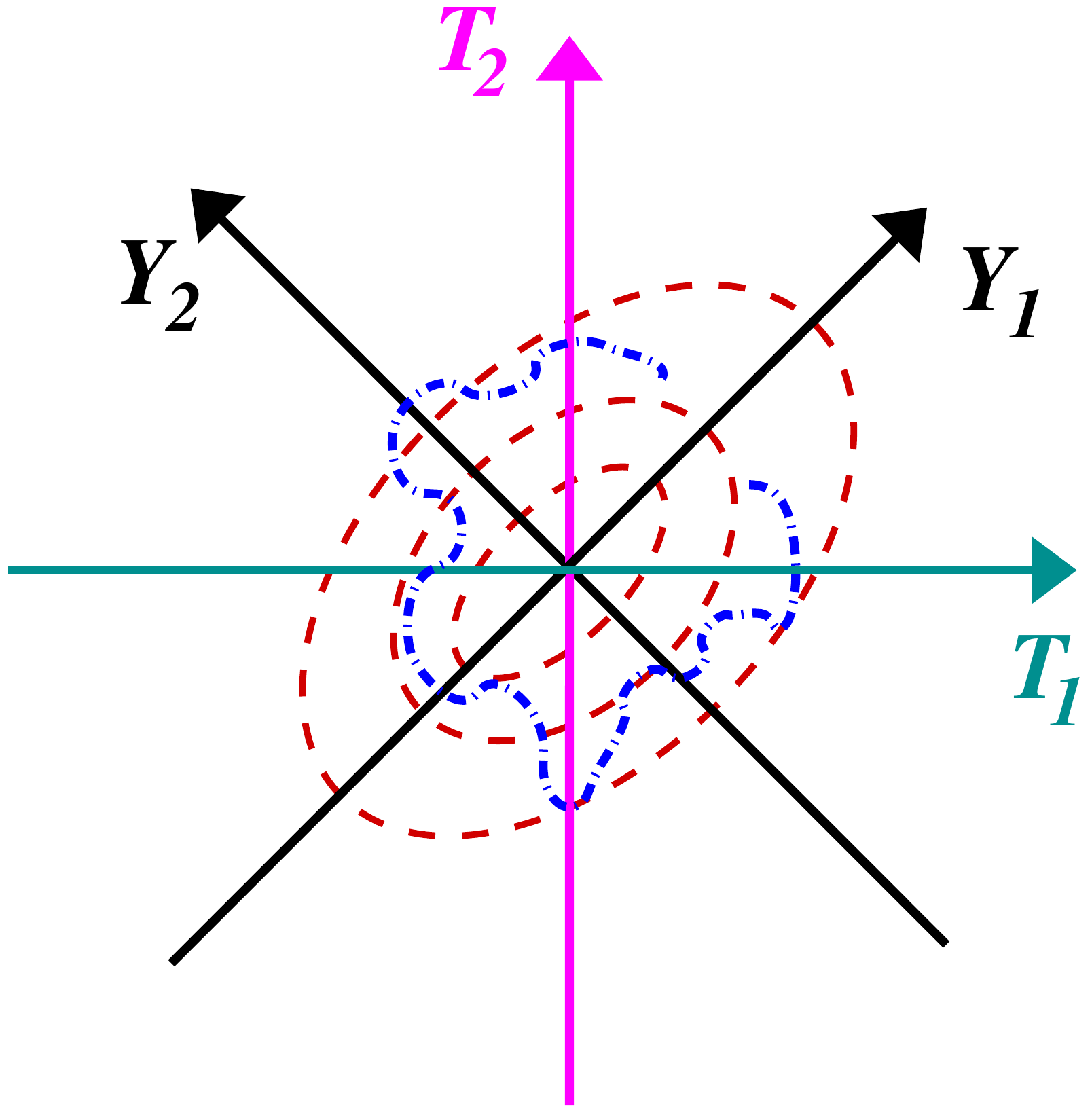}
 \caption{ Brownian motion with distinct thermal reservoirs. 
 Position variable $x_i$ is affected by a bath at temperature $T_i$. 
 The harmonic potential presents principal axes $Y_i$. }
\label{2DLan}
\end{figure}

We consider an overdamped Brownian particle moving in a  two-dimensional space. The system is described by two degrees of freedom, 
$x_1$ and $x_2$, and a potential field given by
%
\begin{equation} \label{QuaPot}
 \begin{split}
  V \left( \mathbf{x}\right)&= \frac{k}{2} \mathbf{x}^{\text{T}} 
   \begin{pmatrix}
   1 & u \\
   u & 1
 \end{pmatrix}  
    \mathbf{x},
 \end{split}
\end{equation}
%
%
where $\mathbf{x}=\left(x_1,x_2\right)$, $\mathbf{x}^{\text{T}}$ is the transpose of $\mathbf{x}$, $k>0$ and $u$ are parameters. In order to characterize a harmonic trap, 
we assume that $-1<u<1$, which leads to a potential landscape with global 
minimum at the origin. For non-trivial values of the parameter $u$, the principal 
axes $Y_i$ of the harmonic potential do not coincide with the Cartesian frame $x_i$, 
as represented in Fig. \ref{2DLan}. This possibility of different eigenframes 
is an important ingredient that affects the stationary behavior of the 
model \cite{FilligerReimann2007}.

The dynamical evolution of the position variables is given by the Langevin equations 
\begin{equation} \label{LanSys1}
 \gamma \dot{x}_i + \int_0^t dt^{\prime} K\left( t - t^{\prime}\right) \dot{x}_i(t^{\prime}) = 
 -\partial_i V + f_i,
\end{equation}
where $\gamma$ is a friction coefficient, $K\left( t \right)$ is the dissipative memory kernel, 
and $f_i$ is an stochastic force related to the thermal environment, with temperature $T_i$, acting 
on $x_i$. The initial conditions are $x_i\left( 0 \right) = 0$ and $\dot{x}_i\left( 0 \right) = 0$. The stochastic system described by \eqref{LanSys1} is a prototype of 
Brownian gyrators \cite{FilligerReimann2007,Argunvolpe2017,ChiangChen2017,CerasoliRondoni2018} 
including non-Markovian properties associated with baths.
We assume the $f_i$ to be Gaussian noises, with temporal correlations 
\cite{CugliandoloKurchan2000,PuglisiVillamaina2009,VillamainaVulpiani2009,Villamaina2013}, 
with cumulants given by  (with $k_B=1$)
%
%
\begin{equation} \label{NoiseCumu1}
 \begin{split}
 \left<f_i(t)\right>_c &= 0, \\
 \left<f_i(t)\,f_j(t^{\prime})\right>_c &= T_i\delta_{ij}\left[ 2\gamma \delta\left( t - t^{\prime} \right) 
 + K\left( t - t^{\prime} \right) \right], \\
  K\left( t \right) &= \frac{\Gamma}{\tau}\exp\left( -\frac{\lvert t \rvert}{\tau} \right),
\end{split}
\end{equation}
where $T_i$ is the effective temperature of bath $i$, with $i=1,2$. $\Gamma$ is a 
friction coefficient, $\tau$ is the correlation time-scale, and $\delta_{ij}$ is the Kronecker delta. 
We assume that the noises are independent, for simplicity sake. Langevin forces described by \eqref{NoiseCumu1} present distinct time-scales effects associated with thermal baths, which have been used to explore physical systems with fast and slow stochastic dynamics \cite{CugliandoloKurchan2000,PuglisiVillamaina2009,VillamainaVulpiani2009,Villamaina2013}. 

For our Brownian particle in a harmonic trap, each non-Markovian bath is 
coupled to a specific position variable, but thermal fluctuations 
affect both degrees of freedom due to the interaction $V(x_1,x_2)$. 
The limiting case of vanishing coupling parameter, $u \to 0$, 
leads to the  $x_i$ evolving independently of each other, under the influence 
of a single reservoir. As a result, if bath temperatures are equal, 
we find equilibrium states characterized by a canonical distribution  
\cite{PuglisiVillamaina2009,Villamaina2013,NascimentoMorgado2019}. Otherwise, for non-zero 
values of $u$ and distinct effective temperatures, we expect the system to
achieve a long-time behavior with rich out-of-equilibrium properties.

We investigate the two-temperature Brownian system \eqref{LanSys1}
through time-averaging procedures \cite{SoaresMorgado2006,SoaresMorgado2008,NascimentoMorgado2019,NascimentoMorgado2020}. 
These techniques are very appropriate to deal with linear Langevin dynamics with general types 
of noises \cite{MorgadoSoares2011,MedeirosQueiros2015,Queiros2016}, and even systems 
with non-linear force fields \cite{MorgadoQueiroz2014,MorgadoQueiroz2016,DefaveriDuarte2017}. 
The main idea is to study the temporal evolution of average quantities such as 
moments and/or cumulants by means of integral transformations. 
Here, we focus on the cumulants of the degrees of freedom $x_i$. 
The main point is that  taking averages related to $x_i$ correspond to taking averages of quantities associated with thermal noises. Technical details are 
shown in Appendix \ref{AppLan}. 
This approach allows us to study many distinct stationary properties of the system, 
such as the distribution function and the average heat exchanged with reservoirs. 

In the next section, we investigate the long-time behavior of the joint 
probability density related to position variables.

%
\section{Probability density} \label{StatProb}

We are interested in determining the distribution associated with 
$x_i$, in order to characterize the stochastic behavior of the system. 
Along the lines of the formalism we are adopting, it is interesting 
to write the instantaneous distribution function as \cite{NascimentoMorgado2019,NascimentoMorgado2020}
\begin{equation} \label{DistFunc0}
 \begin{split}
 P \left( \mathbf{x},t \right) &= \left< \delta \left( \mathbf{x} -
  \mathbf{x}\left( t \right) \right)\right>,
 \end{split}
\end{equation}
where the averages $\left< \cdot \right>$ are taken over the noise realizations. 
Then, we express the delta functions as Fourier integrals,
\begin{equation}
    \delta\left( \mathbf{x} -
  \mathbf{x}\left( t \right) \right) = 
    \mathlarger{\int} \frac{d^2\mathbf{z}}{4\pi^2}  \exp\left[ i\mathbf{z}^{\text{T}}\left( \mathbf{x} -
  \mathbf{x}\left( t \right) \right) \right].
\end{equation}
This gives us the probability density in terms of a Fourier transform of the characteristic function. 
However, since we 
defined the stochastic forces in terms of noise cumulants, it is more appropriate to deal with the 
cumulant-generating function, which is obtaining taking the logarithm of characteristic function. 
As a result, we find
\begin{equation} \label{DistFunc1}
 \begin{split}
 P \left( \mathbf{x},t \right) &= \mathlarger{\int}   \frac{d^2\mathbf{z}}{4\pi^{2}}  
  \exp \left[  i \mathbf{z}^{\text{T}}\mathbf{x} + \mathcal{K} \left( \mathbf{z},t \right) \right],
 \end{split}
\end{equation}
where
%
%
%
\begin{equation} \label{CumuFunc1}
 \begin{split}
   \mathcal{K} \left( \mathbf{z}, t \right) &= -\frac{1}{2} 
   \mathbf{z}^{\text{T}}
   \begin{pmatrix}
    \left< x_1^2 \left( t \right) \right>_c &  \left< x_1 \left( t \right)x_2 \left( t \right) \right>_c \\
     \left< x_1 \left( t \right)x_2 \left( t \right) \right>_c &  \left< x_2^2 \left( t \right) \right>_c
   \end{pmatrix}
   \mathbf{z}, 
       \end{split}
\end{equation} 
is the instantaneous join cumulant-generating function.
However, we may write the cumulants of each $x_i$ 
as a linear superposition of the 
cumulants of stochastic forces.
Then, we can use these cumulants 
to obtain \eqref{CumuFunc1} and, consequently, 
an analytic expression for the join probability density.

\subsection{Stationary cumulants} \label{StatioCumu}\
The harmonic potential as well as the Gaussian nature of the noises lead to a full investigation 
about the instantaneous cumulants. These quantities present many transient effects, which are 
irrelevant as the system achieves stationary states, which  is the regime we are interested in. 
Then, for the long-time limit $t \to \infty$, we can write steady-states cumulants as
\begin{equation} 
b_{ij} = \lim_{t \to \infty} \left< x_i\left( t \right)x_j\left( t \right) \right>_c.
\end{equation}
Technically, these stationary cumulants are obtained by considering 
contour integration methods, as discussed 
in Appendix \ref{AppVar}. Then, alter all calculations, we have
%
%
%
\begin{widetext}
\begin{equation} \label{StatCumu11}
 \begin{split}
  b_{11} &= \frac{  \left[ \zeta \left( 2 -u^2 \right) + 
  k\tau\Gamma\left( 4-3u^2 \right) \right] T_1 +  \left( \zeta + k\tau\Gamma \right) u^2T_2 }{2k\left( 1 -u^2 \right)
  \left[ \zeta + k\tau\Gamma\left(2-u^2 \right)  \right]},
 \end{split}
\end{equation}
\begin{equation} \label{StatCumu22}
 \begin{split}   
  b_{22} &= \frac{ \left( \zeta + k\tau\Gamma \right) u^2T_1 +  \left[ \zeta \left( 2 -u^2 \right) + 
  k\tau\Gamma\left( 4-3u^2 \right) \right] T_2 }{2k\left( 1 -u^2 \right)
  \left[ \zeta + k\tau\Gamma\left(2-u^2 \right)  \right]}, \end{split}
\end{equation}
\end{widetext}
%
%
\begin{equation} \label{StatCumu12}
 \begin{split}
  b_{12} &= -\frac{u\left(T_1 + T_2\right)}{2k\left( 1 -u^2 \right)},\\
 \end{split}
\end{equation}
where 
\begin{equation}
 \zeta = \left( \gamma + \Gamma \right)^2 + k\tau\left[  2\gamma + k\tau\left( 1-u^2 \right) \right],
\end{equation}
is a positive quantity. It is clear that 
these cumulants contain all information about the 
non-Markovian effects on the fluctuations associated with position variables. 
Notice that $b_{11}$ and $b_{22}$ are positive, and $b_{12}$ is negative.

\subsection{Stationary distribution} \label{StatDisP}

Since we calculated all cumulants for stationary states, 
we can express the steady-state join distribution as
%
%
\begin{equation} \label{StatProbDist}
 \begin{split}
  P_s\left( \mathbf{x} \right) &= \frac{1}{2\pi\sqrt{\text{det}\mathbb{B}}} 
  \exp\left( -\frac{1}{2} \mathbf{x}^{\text{T}} \mathbb{B}^{-1}\mathbf{x} \right),
 \end{split}
\end{equation}
%
%
where
%
\begin{equation}
\begin{split}
\mathbb{B} = \begin{pmatrix}
   b_{11} & b_{12} \\
   b_{12} & b_{22}
 \end{pmatrix},
\end{split}
\end{equation}
and $\mathbb{B}^{-1}$ and $\text{det}\mathbb{B}$ are the inverse and the determinant 
of $\mathbb{B}$, respectively. 
The general stationary state is given by a non-equilibrium Gaussian distribution. 
Nevertheless, it is straightforward to perceive that equilibrium states are achieved 
whenever the heat reservoirs present the same temperatures, $T_i \to T$,
\begin{equation} \label{StatProbDistGibbs}
 \begin{split}
 P_s \left( \mathbf{x} \right) &\to  \frac{ k\sqrt{1-u^2}}{2\pi T}  
 \exp\left[  -\frac{V\left(  \mathbf{x} \right)}{T}  \right].
 \end{split}
\end{equation}
which is the canonical distribution for 
the two-dimensional linear system defined by the potential \eqref{QuaPot}.

Another important special case is obtained by taking the 
Markovian limit $\tau \to 0$. 
For this case, we have
%
%
\begin{equation} \label{StatCumutau0X11}
 \begin{split}
  b_{11} & \to \frac{ 2T_1 - \left( T_1 - T_2 \right) u^2  }{2k\left( 1 -u^2 \right)}, \\
 \end{split}
\end{equation}
\begin{equation} \label{StatCumutau0X22}
 \begin{split}
  b_{22} & \to \frac{ 2T_2 + \left( T_1 - T_2 \right) u^2 }{2k\left( 1 -u^2 \right)},	
 \end{split}
\end{equation}
\begin{equation} \label{StatCumutau0X12}
 \begin{split}
  b_{12} & \to -\frac{u\left(T_1 + T_2\right)}{2k\left( 1 -u^2 \right)},\\
 \end{split}
\end{equation}
%
These expressions allow us to write a stationary distribution which is 
in agreement with earlier works on Brownian gyrators with thermal white noises 
at different temperatures \cite{DotsenkoOshanin2013,MancoisWilkowski2018,NascimentoMorgado2020}. 
That memoryless case may present some interesting properties due to the mismatch 
between the principal axes associated with the distribution function and the underlying 
harmonic potential \cite{DotsenkoOshanin2013,MancoisWilkowski2018}. For stationary states, 
the system can exhibit a complex probability flux and non-zero mean angular velocity. 
One can say that non-equilibrium fluctuations induce stochastic rotation as the bath 
temperatures are different \cite{VolpePetrov2006,FilligerReimann2007,DotsenkoOshanin2013,
MancoisWilkowski2018,Argunvolpe2017,ChiangChen2017}. 

Although we have a Brownian system with time-correlated thermal noises, it is 
possible to check that the potential energy \eqref{QuaPot} and the probability 
function \eqref{StatProbDist} present different principal axes. 
This mismatch condition can also lead to effective rotational behavior. 
Then, it is worth investigating how non-Markovian baths can influence the 
induced stochastic rotation of the system.

In the next section, we analyze the effects of memory on 
the average rotational quantities.

\section{Stochastic gyration properties} \label{StoTor}
We want to study the effects of non-Markovian baths on the average 
rotational properties of the system. This is possible by considering 
the torque due to all forces acting on the Brownian particle. 
It is important to mention that earlier theoretical studies describe Brownian 
gyrators with solely thermal white noises 
\cite{VolpePetrov2006,FilligerReimann2007,DotsenkoOshanin2013,
MancoisWilkowski2018,Argunvolpe2017,ChiangChen2017}. 
In this work with non-Markovian reservoirs, one can shows that, by evaluating the torques $\mathcal{T}$ about the origin of all forces, 
the overdamped Langevin system \eqref{LanSys1} leads to a rotational equation of motion of the type 
\begin{equation} \label{BToque1}
 \mathcal{T}_{diss} + \mathcal{T}_{V} + \mathcal{T}_f = 0, 
\end{equation}
where the torques are given by
%
\begin{equation} \label{DissTorque}
 \begin{split}
 \mathcal{T}_{diss} &= -\gamma \varepsilon_{ij}\,x_i\,\dot{x}_j - \mathlarger{\int}_{0}^t dt^{\prime} K(t-t^{\prime}) \,
 \varepsilon_{ij}\,x_i\left( t \right)\,\dot{x}_j\left( t^{\prime} \right),
 \end{split} 
\end{equation}
%
%
\begin{equation} \label{UTorque}
 \begin{split}
 \mathcal{T}_{V} &= - \varepsilon_{ij}\,x_i\,\partial_j V, 
\end{split}
\end{equation}
\begin{equation} \label{FlucTorque}
 \begin{split}
 \mathcal{T}_{f} &= \varepsilon_{ij}\,x_i\,f_j, 
\end{split}
\end{equation}
%
%
and $\varepsilon_{ij}$ is a two-index Levi-Civita symbol, proper for a  two dimensional geometry. We also assume Einstein summation convention. 
The presence of a memory kernel for dissipation implies that 
$\mathcal{T}_{diss}$ should be interpreted as the torque due to 
friction forces. Otherwise, $\mathcal{T}_{V}$ and $\mathcal{T}_{f}$ 
are the torques exerted by harmonic and thermal fluctuating forces, respectively. 
 
It is possible to identity, through relationships between angular momentum 
and angular velocity, a quantity called weighted angular velocity \cite{Argunvolpe2017}, 
which is the angular velocity (about the origin) multiplied by the squared distance 
form the origin. The weighted angular velocity is 
closely related to the dissipative torque $\mathcal{T}_{diss}$. 
The average of $\mathcal{T}_{diss}$ may be evaluated by means of the noise averages and 
integral transformations, as we did for the stationary cumulants. 
If we apply similar techniques to find the average angular velocity \cite{DotsenkoOshanin2013,MancoisWilkowski2018,ChiangChen2017}, 
probably we would obtain a complicated mathematical structure. 
Also, for our model system with memory kernel, 
a direct study of $\mathcal{T}_{diss}$ would lead to cumbersome expressions. 
Nevertheless, the balance equation \eqref{BToque1} indicates that the torque due to friction 
can be estimated by using the torques produced by harmonic and thermal forces. 
These quantities are feasible to calculate and analyze as follows.

\subsection{Average torque due to harmonic force} \label{TorqueHar}

We obtain analytic expressions for all average torques by using 
time-averaging treatments. In particular, harmonic forces 
contribute to the average rotation through the stationary state relation
\begin{equation}
 \left< \varepsilon_{ij}\,x_i\,\partial_j V \right>_s = 
 ku\left( \left< x_1^2 \right>_c - \left< x_2^2 \right>_c \right),
\end{equation}
which is related to the variances of position variables. The subscript $s$ stands for the long-time limit.  
Since we already calculated those quantities for stationary states in \ref{StatioCumu}, we may write
\begin{equation} \label{AveUtoque}
 \begin{split}
 \left< \mathcal{T}_V \right>_s &=
 \frac{u\left(T_2 - T_1\right)\left( \zeta + 2k\tau\Gamma \right)}
  { \zeta + k\tau\Gamma\left(2-u^2 \right) },
 \end{split}
\end{equation}
where $\zeta$ is given by \eqref{ZetaPara}.
The average torque \eqref{AveUtoque} depends on the difference between the 
temperatures of thermal baths. The same is true for the others average torques. 
This dependence indicates the importance of non-equilibrium thermal fluctuations 
on the stochastic rotation. We also notice a non-linear dependence on the coupling 
parameter $u$, in addition to contributions coming 
from noise temporal correlations.

It is worth mentioning that the simplest case of Brownian gyrator with thermal white noises 
exhibits a linear dependence on the coupling parameter $u$. In fact, by taking the Markovian limit of the average harmonic torque, we find
\begin{equation} \label{AveUtoque1}
 \left< \mathcal{T}_V \right>_s \to u\left(T_2 - T_1\right) 
 \quad \text{as} \quad \Gamma,\tau \to 0,
\end{equation}
This is consistent with the memoryless case of an overdamped Brownian system, at 
different effective temperatures, in a two-dimensional confining potential 
\cite{FilligerReimann2007,Argunvolpe2017}.

\subsection{Average torque due to thermal noises} \label{TorqueThermal}

It is reasonable that non-Markovian baths affect the instantaneous behavior of all stochastic 
torques. Nevertheless, we find that thermal fluctuating forces 
also contribute non-trivially to the steady-state rotational properties. 
As shown in Appendix \ref{AppStochTorque}, thermal noises produce an average torque of the type
\begin{equation} \label{AvFlucTorque}
 \left< \mathcal{T}_f \right>_s = 
  \frac{4\Gamma\tau ku\left( T_1 - T_2 \right)}
  {\left[ 2\left( \gamma + k\tau \right) + \Gamma \right]^2 - 4\left( k\tau u \right)^2 }.
\end{equation}
Notice the dependence on friction constant $\Gamma$ and persistence $\tau$. 
The average thermal torque \eqref{AvFlucTorque} is nontrivial as $\Gamma$ and $\tau$ are non-zero. This 
is very different from the Markovian limiting case, which leads 
to a null average thermal torque \cite{FilligerReimann2007,Argunvolpe2017},
\begin{equation}
 \left< \mathcal{T}_f \right>_s \to 0 \quad \text{as}  \quad \Gamma,\tau \to 0.
\end{equation}
The average torque exerted by friction forces is directly related 
to the average torque produced by harmonic forces if memory effects are vanishing. 
However, for the general non-Markovian case, both dissipative and fluctuating forces may lead 
to relevant contributions to the average rotation.

Therefore, it seems that Brownian gyrators with friction memory kernel 
may exhibit systematic rotational dynamics distinct from the Markovian case. 
Noise temporal correlations can also influence the energetic 
balance and heat exchanges with thermal environments. 
Then, it is important to develop a detailed analysis about 
the energetic considerations of the model as well as its irreversibility properties.

In the next section, we study how memory may affect the 
behavior of the stochastic energetics of the system.

\section{ Steady-state heat exchange } \label{StoEner}
We can also investigate the stochastic behavior of the heat flow 
and energetic variations for the our two-temperature Langevin dynamics. 
In order to do that, we follow the stochastic thermodynamics formalism 
discussed by Sekimoto\cite{Sekimoto1998,Sekimoto2010}. For this kind of 
formalism, it is more appropriate to adopt calculus manipulation in 
Stratonovich sense \cite{Gardiner,livro_vankampen}. 

Based on the overdamped Langevin equations \eqref{LanSys1}, 
which present trivial external work contributions
\begin{equation} \label{BalanceEnergy}
    d^{\prime}\mathcal{W}= dV - d^{\prime}\mathcal{Q}_1 - 
    d^{\prime}\mathcal{Q}_2= 0,
\end{equation}
we can write the stochastic heat exchanged  as
\begin{equation} \label{StochHeatXi}
 d^{\prime}\mathcal{Q}_i  = \,\partial_i V dx_i = J_i\,dt, 
\end{equation} 
%
where
\begin{equation} \label{HeatJ}
 \begin{split}
 J_i &= f_{i}\,\dot{x}_i - \gamma\,\dot{x}_i^2 -\mathlarger{\int}_{0}^t dt^{\prime} K\left( t - t^{\prime} \right) 
 \dot{x}_i\left( t \right)\dot{x}_i\left( t^{\prime} \right).
 \end{split}
\end{equation}
is the heat flux related to bath $i$, at temperature $T_i$. The stochastic equation \eqref{BalanceEnergy} represents the energetic balance 
for the coordinate $x_i$, and \eqref{HeatJ} is interpreted as the 
power contributions of thermal baths. Noise temporal correlations contribute to 
the stochastic heat by means of the friction memory kernel. 
The heat flow in \eqref{HeatJ} is consistent with the kind of non-Markovian system defined by 
\eqref{LanSys1}. For vanishing contributions of the memory kernel, 
which is corresponds to $\Gamma \to 0$ in \eqref{NoiseCumu1}, we obtain the 
memoryless case
\begin{equation}
 J_i \to f_{i}\,\dot{x}_i - \gamma\,\dot{x}_i^2 \quad \text{as} \quad \Gamma \to 0, 
\end{equation}
which is in agreement with the stochastic heat flux obtained from a 
Brownian system with Markovian thermal baths \citep{Sekimoto1998,Sekimoto2010}.

The balance equation \eqref{BalanceEnergy} relates the 
dissipative and injected power of reservoirs to the internal energy variation. 
The stochastic heat exchanges during a time interval 
between $t$ and $t + t_o$ is given by
\begin{equation} \label{StoHeat12}
 \begin{split}
  \mathcal{Q}_i &= \int_{t}^{t + t_o} dt^{\prime} \dot{x}_i\,\partial_i V.
 \end{split} 
\end{equation}
With these expressions, it is straightforward to write a stochastic version 
of the first law of the 
thermodynamics, which reads 
$\mathcal{Q}_1 + \mathcal{Q}_2 = V \left( t + t_o \right) - V \left( t  \right)$.
%
%
This relation depends on the trajectory along which the system evolves from 
$t$ to $t + t_o$. However, our main interest is to analyze the average properties in the 
steady-state regime. As a result, for the long-time limit, we have
\begin{equation} \label{Av1Law}
 \left< \mathcal{Q}_1 \right>_s + \left< \mathcal{Q}_2 \right>_s  = 0, \quad \text{as} \quad t \to \infty,
\end{equation}
which is a constraint on the average heat flowing from one bath to the other as the 
stationary state is achieved. From a technical perspective, the constraint \eqref{Av1Law} allows us to investigate the heat exchanges through the system in terms of a specific position variable or reservoir. 
Then, for simplicity, we choose $x_1$ (or reservoir $1$) to work with.

\begin{figure}
 \centering
 \includegraphics[scale=0.3]{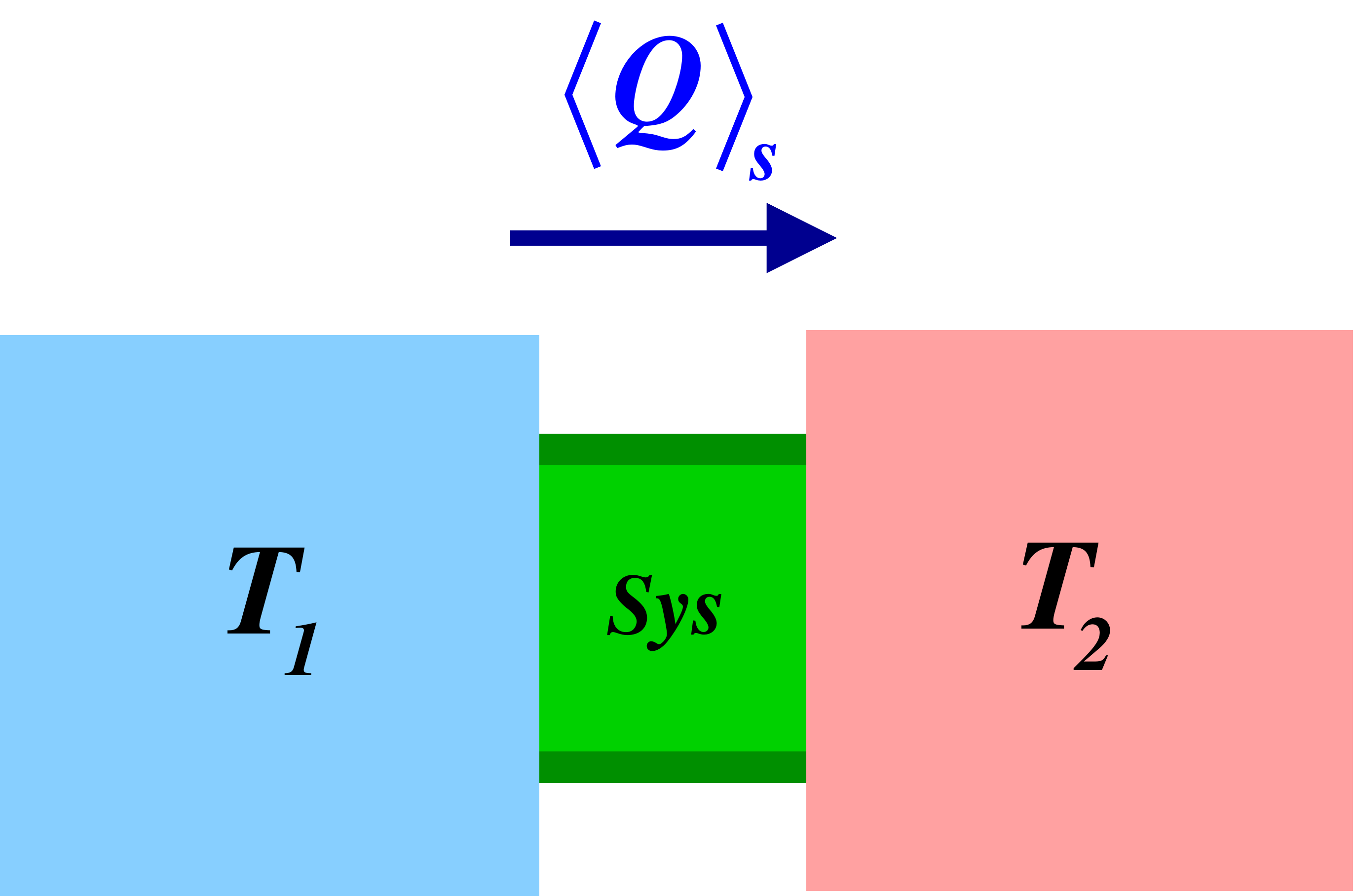}
 \caption{ Stationary average heat flow through the system. 
 For $T_1>T_2$ in long-time limit, the system absorbs, on average,  an amount of  heat from reservoir $1$ and releases the same amount into reservoir $2$. }
\label{2DHeat}
\end{figure}

It is possible to develop many mathematical calculations for the linear 
model we are studying, specially for the steady-state stochastic energetics 
(technical details shown in Appendix \ref{AppStochEner}). Then, by considering the 
long-time limit $t \to \infty$, we find the stationary average heat
%
\begin{equation} \label{AveStotHeat1}
 \begin{split}
  \left< \mathcal{Q}_1 \right>_s &= \frac{\left( T_1 - T_2 \right)t_o k u^2\Psi}{2\gamma} 
   \end{split}
\end{equation} 
%
with
\begin{equation} \label{AveStotHeat11}
 \begin{split}
  \Psi &= \frac{ \gamma \left( \gamma + \Gamma \right) + 
  k\tau \left[ 2\gamma + \Gamma +  k\tau \left( 1-u^2 \right)   \right] }
  { \left( \gamma + \Gamma \right)^2 + k\tau \left[ 2\gamma + \Gamma + 
  \left( \Gamma + k\tau \right) \left( 1-u^2 \right) \right] }.
 \end{split}
\end{equation} 
This is valid for any values of $u^2<1$. 
Notice the positive quantity $\Psi$ is less than unit.
It is important to bear in mind that the system is simultaneously in  contact  with thermal baths at different temperatures. 
As a result, for $T_1>T_2$, average heat (related to a time interval $t_o$) is absorbed from reservoir $1$ and delivered into reservoir $2$, as depicted in Fig. \ref{2DHeat}.  
In fact, the expression \eqref{AveStotHeat1} is consistent with the interpretation that, for heat flowing into 
the system through $x_1$, the average heat exchange $\left< \mathcal{Q}_1 \right>_s$ should 
be positive whenever $T_1>T_2$. Another point is that \eqref{AveStotHeat1} exhibits a very 
complicated dependence on the noise temporal correlations as well as the interaction parameter $u$.

%

Clearly, the average heat \eqref{AveStotHeat1} may be null 
by taking equal bath temperatures, provided that $u$ is non-zero, 
for any finite values of friction constants and persistence time-scale. 
For the particular case of Markovian thermal baths, it is possible to recover 
results consistent with the memoryless Brownian gyrator. 
In fact, this can be done by taking the limit $\Gamma \to 0 $ in \eqref{AveStotHeat11}, which leads to $\Psi \to 1$. Then, we find
\begin{equation}
 \left< \mathcal{Q}_1 \right>_s \to 
 \frac{ \left( T_1 - T_2 \right)t_o k u^2}{2\gamma},
\end{equation}
which is in agreement with the average heat obtained 
from a two-temperature Langevin system with Gaussian white noises 
\cite{Sekimoto1998,Sekimoto2010}.
Notice that by taking the limit $\tau \to 0$ in \eqref{AveStotHeat11}, we find an average heat with friction coefficient $\gamma + \Gamma$.

As each degree of freedom is affected by a single bath with 
a specific temperature $T_i$, the whole system experiences an irreversible heat exchange with reservoirs, provided that $T_1 \neq T_2$. In the next section, we consider the entropic changes of the model.

\section{Stationary entropy variation} \label{StochEntroVar}

The irreversibility aspects of our Langevin system is characterized by determining 
the total entropy generation for the long-term run. In order to do that, we follow 
the approach developed previously by the authors~\cite{NascimentoMorgado2020}, 
which studied a similar two-temperature Brownian system with friction memory kernel. 
Accordingly, we assume that, during the time interval between $t$ and $t + t_o$, 
the total entropy change of 
the the system and the medium is given by
\begin{equation} \label{TotEntro}
 \Delta S = \Delta S_{sys} + \Delta S_{m},
\end{equation}
where $\Delta S_{sys}$  is the change of entropy associated with 
the system, which is the Brownian particle subjected to the harmonic potential, 
and $\Delta S_{m}$ is the entropy variation related to the thermal medium, which are the 
non-Markovian heat reservoirs.

We define, as usual, the entropy of the system in term of the well-known 
Gibbs-Shannon entropy formula \cite{Tome2006,TomeOliveira2015} (we assume Boltzmann constant $k_B=1$)
\begin{equation}
 S_{sys} \left(t \right) = -\int d^2 \mathbf{x} P\left( \mathbf{x},t \right)
 \ln P\left( \mathbf{x},t \right),
\end{equation}
where $P\left( x_1, x_2,t\right)$ is the time-dependence join 
distribution \eqref{DistFunc1}. 
Clearly, the entropy variation of the system is of the type $ \Delta S_{sys} = S_{sys} \left( t + t_o \right) - S_{sys} \left( t \right)$.
%
%
The entropy change of the thermal medium is associated with the 
stochastic heat exchanges relate to thermal reservoirs \cite{Seifert2012}. 
Since we are interested in the average properties, we can assume 
\begin{equation} \label{VarEntroMed}
 \begin{split}
 \Delta S_{m} &= -\frac{\left< \mathcal{Q}_1 \right>}{T_1}
 -\frac{\left< \mathcal{Q}_2 \right>}{T_2}, 
 \end{split}
\end{equation}
which is the average version of entropy variation 
associated with the stochastic heat flowing through the thermal environment.
It is worth mentioning that the definition of stochastic heat in \eqref{HeatJ} 
indicates that a positive heat quantity accounts to increase the internal energy of the system. 

The main interest here is to determine the stationary behavior of the total entropy variation. 
However, as the system achieves the steady-state regime, the probability 
distribution becomes time-independent. 
As a result, for the long-term run, it straightforward to perceive that 
the entropy change of the system is typically null, since $\Delta S_{sys} \to 0$ as $t \to \infty$.
%
%
This means that, for stationary states, there exists an entropic balance 
between internal system and surroundings. 
On the other hand, the entropy change of the thermal medium is related 
to the steady-state heat exchanges, which gives 
\begin{equation} \label{VarEntroMed1}
 \begin{split}
 \Delta S_{m}^s &= \frac{T_1 - T_2}{T_1T_2}\left< \mathcal{Q}_1 \right>_s.
 \end{split}
\end{equation}
Here, we use the first law form in \eqref{Av1Law}, which shows that 
thermal energy injected by one reservoir 
is, on average, dissipated into the other.

Therefore, combining all relevant entropic contributions for the 
stationary states, we obtain
\begin{equation} 
 \begin{split}
 \Delta S_{tot} \to \Delta S_{m}^s &= \sigma t_o \quad \text{as} \quad t \to \infty,
 \end{split}
\end{equation}
where
%
%
\begin{equation} \label{EntroProd}
 \begin{split}
 \sigma &= \frac{\left( T_1 - T_2 \right)^2 ku^2 \Psi}{2\gamma T_1 T_2},
  \end{split}
\end{equation}
%
is the total steady-state entropy production rate, which is clearly 
a non-negative quantity. The parameter $\Psi$ given by \eqref{AveStotHeat11} is a positive quantity. Notice that the persistence parameter $\tau$ and the coupling parameter $u$ affect the entropy generation. In fact, the non-Markovian properties of the baths 
play a role in the heat flow and, consequently, the steady-state entropy behavior of the system. 
We can obtain results in agreement with equilibrium statistical 
thermodynamics for non-zero values of $u$ and equal bath temperatures.  In this case, microscopic reversibility is recovered and stationary states are characterized by a Boltzmann-Gibbs distribution, as 
shown in \ref{StatDisP}.

The rate of entropy production \eqref{EntroProd} is consistent with the particular 
case of a Brownian dynamics with white noises and distinct temperatures. 
This limiting case is obtained straightforwardly by considering 
$\Gamma \to 0$ in \eqref{EntroProd}. Consequently, we find
\begin{equation} \label{EntroProdGamma=0}
 \begin{split}
 \sigma \to \frac{\left( T_1 - T_2 \right)^2 ku^2}{2\gamma T_1 T_2} 
 \quad \text{as} \quad \Gamma \to 0.
 \end{split}
\end{equation}
This entropy generation rate is due to a non-zero average heat 
flowing through the system, which leads to out-of-equilibrium stationary states, 
stressing the role heat reservoirs at distinct temperatures play.

Earlier works also discuss the 
effects of memory on the irreversibility associated with Langevin systems 
with many thermal baths \cite{CugliandoloKurchan2000,PuglisiVillamaina2009}. 
Our model presents qualitative results analogous to those investigations. 
For example, we find that the memory effect contributes to the stationary entropy generation, 
but these contributions are trivial as the bath temperatures become equal, 
even for finite values of noise temporal correlations $\tau$. This is also observed in a one-dimensional 
Brownian system coupled to many heat reservoirs \cite{PuglisiVillamaina2009,Villamaina2013}.

Langevin dynamics with multiple reservoirs present intriguing physical properties. 
For simple linear models analogous to the one studied in this work, we believe it is 
important to develop futher investigations with the inclusion of inertia, 
distinct kinds of non-Markovian noises, and athermal reservoirs. Also, it would be very interesting to 
consider model systems with non-linear force fields, where it is possible to identify 
some noise-induced effects \cite{Spagnolo2009,Spagnolo2015,Spagnolo2017}. 

\section{Conclusions} \label{Con}
We study an elementary Brownian gyrator with non-Markovian thermal baths. 
The joint probability density associated 
with position variables is calculated analytically by means of a time-averaging formalism. 
%
%
We find that memory affects the stochastic rotational 
properties in the long-time limit. 
For a finite memory time-scale, we show there exists non-trivial 
average torques due to friction forces, harmonic and fluctuating thermal forces. 
Nevertheless. For the memoryless limit, we observe a null average torque due to 
thermal noises. 
%
The stochastic energetic properties of the model is investigated for steady-states. 
The average heat and the entropy production show a memory-dependent behavior. 
As the baths present different temperatures, memory contributes 
to the long-time behavior of the entropy generation. 
Otherwise, for reservoirs with equal temperatures, the entropy production is null, 
even for finite memory time-scale, and equilibrium is recovered.

\section*{Acknowledgment}
This work is supported by the Brazilian funding agencies CNPq and CAPES (Finance Code 001).
\appendix
%
%
\section{Frequency-domain representation of Langevin equations} \label{AppLan}

Following \cite{NascimentoMorgado2019,NascimentoMorgado2020}, we introduce the integral form 
\begin{equation} \label{LaplaceInt}
 x_j\left(t\right) = \lim_{\epsilon \to 0}
 \mathlarger{\int} \frac{dq}{2\pi}  e^{\left( iq + \epsilon \right)t}\widehat{x}_j
 \left( iq + \epsilon  \right),
\end{equation}
where $\widehat{x}_j\left(s \right)$ is is the Laplace transform of $x_j\left(t \right)$. 
%
%
Now, we take the Laplace transform of the Langevin equations, which gives 
%
%
%
\begin{equation} \label{LapX1X2A}
 \begin{split} 
  \widehat{x}_1  &= \frac{1+\tau s}{p}\left\{ \left[ \gamma\tau s^2 + \left( \gamma + \Gamma + k\tau \right)s +
  k \right]\widehat{f}_1  -ku\left( 1 + \tau s \right)\widehat{f}_2 \right\},  
  \end{split} 
\end{equation}
\begin{equation} \label{LapX1X2B}
 \begin{split} 
   \widehat{x}_2 &= \frac{1+\tau s}{p}\left\{ -ku\left( 1 + \tau s \right)\widehat{f}_1 + \left[ \gamma\tau s^2 + \left( \gamma + \Gamma + k\tau  \right)s + k \right] \widehat{f}_2 \right\},
  \end{split} 
\end{equation}
where
\begin{equation} \label{Req}
 \begin{split} 
  p &= p_{+}p_{-}, \\
  p_{\pm} &= \gamma\tau s^2 + \left[ \gamma + \Gamma + k\tau\left( 1 \pm u \right)  \right]s 
  + k\left( 1 \pm u \right).
 \end{split} 
\end{equation}
%
%
We also need the Laplace transforms of noise cumulants 
%
%
\begin{equation} \label{LapNoiseCumu1}
 \begin{split}
 \left<\widehat{f}_i(s_1)\,\widehat{f}_j(s_2)\right>_c &= 
 \frac{T_i\delta_{ij}}{{s_1 + s_2}} \bigg\{ 2\gamma + \frac{\left[2 + \tau\left( s_1 + s_2 \right) \right]\Gamma}
 {\left( 1 + \tau s_1 \right)\left( 1 + \tau s_2 \right)} \bigg\}.
 \end{split}
\end{equation}
%
%
Therefore, in frequency-domain representation, the cumulants related to position variables are 
written as linear combination of noises cumulants
%
%
\section{Variances for steady-states} \label{AppVar}
We consider the change of variables
\begin{equation} \label{TranfNew}
 \begin{split}
  y_1 & = x_1 + x_2, \\
  y_2 & = x_1 - x_2.
 \end{split}
\end{equation}
%
%
%
%
%
Then, we use the integral relation \eqref{LaplaceInt} and the Laplace forms of noise cumulants \eqref{LapNoiseCumu1}, which give $\left< y_iy_j \right>_c$ as 
integral expressions that can be calculated applying the methods of residues. 
Then, for the long-time limit, one may find
%
%
%
\begin{equation}
 \begin{split}
  \lim_{t \to \infty}\left< y_1^2\left( t \right) \right>_c &= \frac{T_1 + T_2}{k\left( 1 + u \right)},
 \end{split}
\end{equation}
\begin{equation}
  \begin{split}
  \lim_{t \to \infty}\left< y_1\left( t \right)y_2\left( t \right) \right>_c &= 
  \frac{\left(T_1 - T_2\right) \left( \zeta + 2k\tau\Gamma \right)}
  { k\left[ \zeta + k\tau\Gamma\left(2-u^2 \right) \right] },
 \end{split}
\end{equation}
\begin{equation}
 \begin{split}
  \lim_{t \to \infty}\left< y_2^2\left( t \right) \right>_c &= \frac{T_1 + T_2}{k\left( 1 - u \right)},
 \end{split}
\end{equation}
%
where
\begin{equation} \label{ZetaPara}
 \zeta = \left( \gamma + \Gamma \right)^2 + k\tau\left[  2\gamma + k\tau\left( 1-u^2 \right) \right].
\end{equation}
%
%
%
%
These expressions allow us to find the stationary cumulants of $x_i$ by taking the 
inverse transform of \eqref{TranfNew}.

%
%
%
%
\section{ Stochastic torque due to thermal noises} \label{AppStochTorque}

The average torque exerted by fluctuating thermal forces is given by 
%
%
%
\begin{equation}
 \begin{split}
  \left< \mathcal{T}_f \right> &=  \left< x_1f_2 - x_2f_1 \right>,
 \end{split} 
\end{equation}
%
%
%
%
where can use the integral form \eqref{LaplaceInt} in order 
to determine the average torque for stationary states. 
Then, we have
\begin{equation}
 \begin{split}
  \left< \mathcal{T}_f \right>_s &= \frac{ku\left( T_1 - T_2 \right)}{\pi} \lim_{\epsilon \to 0} 
  \mathlarger{\int} dq_1\frac{ \gamma \left[ 1 - \tau^2\left( iq_1 + \epsilon \right)^2 \right] + \Gamma }
  { p_+\left( iq_1 + \epsilon \right)p_-\left(iq_1 + \epsilon \right) } \left\{ \frac{ 1 + \tau \left(iq_1 + \epsilon \right) }{1 - \tau \left(iq_1 + \epsilon \right)} \right\}.
 \end{split}
\end{equation}
This integral over $q_1$ does not present any convergence problem. 
As a result, by using the methods of residues, we obtain
\begin{equation}
 \left< \mathcal{T}_f \right>_s = 
  \frac{4\Gamma\tau ku\left( T_1 - T_2 \right)}
  { \left[ 2\left( \gamma + k\tau \right)  + \Gamma \right]^2 - 4\left( k\tau \right)^2}.
\end{equation}
%


%
%
%
%
\section{ Average stochastic heat } \label{AppStochEner}
The average heat associated with the thermal bath $1$ is of the form
\begin{equation} \label{HeatQ1}
 \begin{split}
  \left< \mathcal{Q}_1 \right> &= C_1 + C_2 + \frac{ku}{4}\left[\Upsilon_1\left(  t + t_o \right) - \Upsilon_1\left(  t \right) \right]  , 
 \end{split} 
\end{equation}
where
\begin{equation}
 C_1=\frac{k}{2} \left[ \left< x_1^2\left( t + t_o \right) \right> - 
  \left< x_1^2\left( t \right) \right> \right],
\end{equation}
\begin{equation}
 \begin{split}
 C_2 &= \frac{ku}{8} 
  \left[ \left< y_1^2\left( t + t_o \right) \right> - 
  \left< y_1^2\left( t \right) \right> + \left< y_2^2\left( t + t_o \right) \right> - 
  \left< y_2^2\left( t \right) \right> \right],
  \end{split}
\end{equation}
and
\begin{equation} \label{UpsilonInt}
 \Upsilon_1\left(  t \right) = \int_{0}^{t} dt^{\prime}
 \left( \left< y_1\,\dot{y}_2 \right> -  \left< y_2\,\dot{y}_1 \right>   \right).
\end{equation}
%
%
%
Then, in terms of the integral representation \eqref{LaplaceInt}, 
we can evaluate \eqref{UpsilonInt} for the long-term run ($t \gg 1$), which gives
\begin{equation}
 \begin{split}
  \Upsilon_1^s &=- \frac{\left( T_1 - T_2 \right)t}{\pi} \lim_{\epsilon \to 0} 
  \mathlarger{\int} \frac{dq_1\left(iq_1 +\epsilon\right)}
  { p_+\left( iq_1 + \epsilon \right) }  \left\{\frac{\gamma\left[ 1 -   \tau^2\left( iq_1 + \epsilon \right)^2 \right] + 
  \Gamma }{p_-\left( -iq_1 - \epsilon \right)} \right\},
 \end{split}
\end{equation}
This integral over $q_1$ does 
not satisfy the Jordan's lemma, since the integrand is a rational function of 
the type $1/q_1$. Then, we should take into account integration along 
the semi-circular contour, which asymptotic form gives
\begin{equation}
 \frac{4\pi t \left( T_1 - T_2 \right)}
 {\gamma} \quad \text{as} \quad \lvert q_1 \rvert \to \infty.
\end{equation}
After evaluating all residue contributions, it is possible to show that
\begin{equation}
 \begin{split}
  \Upsilon_1^s &= \frac{2 t u \left( T_1 - T_2 \right)}{\gamma} 
  \frac{ \lozenge_1 }{ \lozenge_2 },
 \end{split}
\end{equation}
where
\begin{equation}
 \begin{split}
  \lozenge_1 &=\gamma \left( \gamma + \Gamma \right) + 
  k\tau \left[ 2\gamma + \Gamma +  k\tau \left( 1-u^2 \right) \right],
 \end{split}
\end{equation}
and
\begin{equation}
 \begin{split}
  \lozenge_2 &=\left( \gamma + \Gamma \right)^2 + k\tau 
  \left[ 2\gamma + \Gamma  + \left( \Gamma + k\tau \right) \left( 1-u^2 \right) \right].
 \end{split}
\end{equation}
The cumulants of $x_i$ and $y_i$ are time-independent for stationary states, which results
\begin{equation}
 C_1,C_2 \to 0 \quad \text{as} \quad t \to \infty.
\end{equation}
Therefore, the steady-state form of \eqref{HeatQ1} is given by
%
\begin{equation}
 \begin{split}
  \left< \mathcal{Q}_1 \right>_s &= \frac{t_o k u^2\left( T_1 - T_2 \right)}{2\gamma} 
  \frac{ \lozenge_1 }{ \lozenge_2 }.
 \end{split}
\end{equation}
%
%
%
%

%
%

\end{document}